\documentstyle[12pt]{article}

\let\useblackboard=\iftrue
\iftrue
\fi

\def\hybrid{\topmargin -20pt  \oddsidemargin 0pt
      \headheight 0pt   \headsep 0pt
      \textwidth 6.25in % A4 paper
      \textheight 9.5in % A4 paper
      \marginparwidth .875in
      \parskip 5pt plus 1pt   \jot = 1.5ex}

\hybrid

\let\LARGE=\large

\let\large=\normalsize

\begin{document}
%\titlepage
\def\x{\times}
\def\beq{\begin{equation}}
\def\eeq{\end{equation}}
\def\beqa{\begin{eqnarray}}
\def\eeqa{\end{eqnarray}}
\def\L{ {\cal L}}
\def\C{ {\cal C}}
\def\N{ {\cal N}}
\def\calE{{\cal E}}
\def\lin{{\rm lin}}
\def\Tr{{\rm Tr}}
\def\cF{{\cal F}}
\def\cD{{\cal D}}
\def\modS{{S+\bar S}}
\def\mods{{s+\bar s}}
\newcommand{\Fg}[1]{{F}^{({#1})}}
\newcommand{\cFg}[1]{{\cal F}^{({#1})}}
\newcommand{\cFgc}[1]{{\cal F}^{({#1})\,{\rm cov}}}
\newcommand{\Fgc}[1]{{F}^{({#1})\,{\rm cov}}}
\def\mpl{m_{\rm Planck}}
\def\mxth{\mathsurround=0pt }
\def\xversim#1#2{\lower2.pt\vbox{\baselineskip0pt \lineskip-.5pt
x  \ialign{$\mxth#1\hfil##\hfil$\crcr#2\crcr\sim\crcr}}}
\def\simgr{\mathrel{\mathpalette\xversim >}}
\def\simle{\mathrel{\mathpalette\xversim <}}

\newcommand{\ms}[1]{\mbox{\scriptsize #1}}
\renewcommand{\a}{\alpha}
\renewcommand{\b}{\beta}
\renewcommand{\c}{\gamma}
\renewcommand{\d}{\delta}
\newcommand{\th}{\theta}
\newcommand{\TH}{\Theta}
\newcommand{\pa}{\partial}
\newcommand{\g}{\gamma}
\newcommand{\G}{\Gamma}
\newcommand{\A}{\Alpha}
\newcommand{\B}{\Beta}
\newcommand{\D}{\Delta}
\newcommand{\e}{\epsilon}
\newcommand{\E}{\Epsilon}
\newcommand{\z}{\zeta}
\newcommand{\Z}{\Zeta}
\newcommand{\k}{\kappa}
\newcommand{\K}{\Kappa}
\renewcommand{\l}{\lambda}
\renewcommand{\L}{\Lambda}
\newcommand{\m}{\mu}
\newcommand{\M}{\Mu}
\newcommand{\n}{\nu}
\newcommand{\X}{\Chi}
\newcommand{\R}{\Rho}
\newcommand{\s}{\sigma}
\renewcommand{\S}{\Sigma}
\renewcommand{\t}{\tau}
\newcommand{\T}{\Tau}
\newcommand{\y}{\upsilon}
\newcommand{\Y}{\upsilon}
\renewcommand{\o}{\omega}
\newcommand{\q}{\theta}
\newcommand{\h}{\eta}

\def\dota{ {\dot{\alpha}} }
\def\lag{Lagrangian}
\def\Kahler{K\"{a}hler}
\def\kahler{K\"{a}hler}
\def\A{ {\cal A}}
\def\C{ {\cal C}}
\def\F{{\cal F}}
\def\cL{ {\cal L}}

\def\R{ {\cal R}}
\def\x{ \times }
\def\beq{\begin{equation}}
\def\eeq{\end{equation}}
\def\beqa{\begin{eqnarray}}
\def\eeqa{\end{eqnarray}}

\sloppy
\newcommand{\ba}{\begin{array}}
\newcommand{\ea}{\end{array}}
\newcommand{\eq}{\begin{equation}}

\newcommand{\ov}{\overline}
\newcommand{\un}{\underline}
\newcommand{\p}{\partial}
\newcommand{\la}{\langle}
\newcommand{\ra}{\rangle}
\newcommand{\bl}{\boldmath}
\newcommand{\ds}{\displaystyle}
\newcommand{\nl}{\newline}
\newcommand{\Nzahl}{{\bf N}  }
\newcommand{\zzahl}{ {\bf Z} }
\newcommand{\Zzahl}{ {\bf Z} }
\newcommand{\Qzahl}{ {\bf Q}  }
\newcommand{\Rzahl}{ {\bf R} }
\newcommand{\Czahl}{ {\bf C} }
\newcommand{\wt}{\widetilde}
\newcommand{\wh}{\widehat}
\newcommand{\fs}[1]{\mbox{\scriptsize \bf #1}}
\newcommand{\ft}[1]{\mbox{\tiny \bf #1}}
\newtheorem{satz}{Satz}[section]
\newenvironment{Satz}{\begin{satz} \sf}{\end{satz}}
\newtheorem{definition}{Definition}[section]
\newenvironment{Definition}{\begin{definition} \rm}{\end{definition}}
\newtheorem{bem}{Bemerkung}
\newenvironment{Bem}{\begin{bem} \rm}{\end{bem}}
\newtheorem{bsp}{Beispiel}
\newenvironment{Bsp}{\begin{bsp} \rm}{\end{bsp}}
%\renewcommand{\arraystretch}{0.5}

%\addtocounter{section}{1}

\renewcommand{\thesection}{\arabic{section}}
\renewcommand{\theequation}{\thesection.\arabic{equation}}

%\setcounter{section}{1}
%\addtocounter{section}{1}
\parindent0em

\useblackboard
\typeout{If you do not have msbm (blackboard bold) fonts,}
\typeout{change the option at the top of the tex file.}
\font\blackboard=msbm10 scaled \magstep1
\font\blackboards=msbm7
\font\blackboardss=msbm5
\newfam\black
\textfont\black=\blackboard
\scriptfont\black=\blackboards
\scriptscriptfont\black=\blackboardss
\def\Bbb#1{{\fam\black\relax#1}}
\else
\def\Bbb{\bf}
\fi

%\renewcommand{\para}{\arabic{paragraph}}
%%\renewcommand{\thesubsection}{\arabic{subsection}}
%\def\sub#1{\thesubsection{#1}}
%%\renewcommand{\theequation}{\thesubsection.\arabic{equation}}
%%\renewcommand{\thefootnote}{\fnsymbol{footnote}}
%\def\sect#1{\subsection{#1}\setcounter{equation}{0}}
%citation styles
%\catcode`@=11
%\def\@cite#1#2{\if@tempswa [#1]\else$^{\scriptscriptstyle
%\mbox{\rm\scriptsize#1}}$\fi}
%useful abbrev.

\newcommand{\eqn}{\begin{eqnarray}}
\newcommand{\enq}{\end{eqnarray}}
\newcommand{\eqa}{\begin{array}}
\newcommand{\ena}{\end{array}}
\newcommand{\en}{\end{equation}}
\newcommand{\ie}{{\it i.e.}}
\newcommand{\no}{\nonumber}
\newcommand{\noi}{\noindent}
\newcommand{\ZZ}{Z\!\!\!Z}
\def\kite#1#2{$^{\mbox{\footnotesize \rm   #1-#2 \normalsize}}$}
\def\skite#1{$^{\mbox{\footnotesize \rm   #1 \normalsize}}$}
\def\comments#1{}
\newcommand{\IZ}{{\Bbb{Z}}}
\def\N{N}
\def\1N{$1\over N$}
\def\CC{\Bbb{C}}
\def\RR{\Bbb{R}}
\def\IZ{\Bbb{Z}}
\def\del{\partial}
\def\half{{1\over 2}}
\def\hhalf{{1\over 4}}
\def\Tr{{\rm Tr\ }}
\def\im{{\rm Im\hskip0.1em}}
\def\bra#1{{\langle}#1|}
\def\ket#1{|#1\rangle}
\def\vev#1{\langle{#1}\rangle}
\def\CT{\cal T}
\def\Dslash{\rlap{\hskip0.2em/}D}

%\setlength{\unitlength}{0.25cm}

%%%%%%%%%%%%%%%%%%%%%%%%%%%%%%%%%%%%%%%%%%%%%%%%%%%%%%

\begin{titlepage}
\begin{center}
\hfill CERN-TH/97-221\\
\hfill {\tt hep-th/9709174}\\

\vskip 3cm

{ \LARGE \bf Finite Energy Solutions in Three-Dimensional \\ 
 Heterotic String Theory}

\vskip .3in

{\bf Mich\`ele Bourdeau and 
Gabriel Lopes Cardoso
}\footnote{\mbox{Email: \tt 
bourdeau@mail.cern.ch, cardoso@mail.cern.ch}}
\\
%\vskip 1.2cm
\vskip 1cm

{\em Theory Division, CERN, CH-1211 Geneva 23, Switzerland}\\

\vskip .1in

\end{center}

\vskip .2in

\begin{center} {\bf ABSTRACT } \end{center}
\begin{quotation}\noi
We show that a large class of supersymmetric solutions to 
the low-energy effective field theory of heterotic string theory
compactified on a seven torus
can have finite energy, which we compute.
The mechanism by which these solutions are turned into finite
energy solutions is similar to the one occurring in the context
of four-dimensional stringy cosmic string solutions.
We also describe the solutions in terms of intersecting eleven-dimensional
M-branes, M-waves and M-monopoles.

\end{quotation}
\vskip 5cm
CERN-TH/97-221\\
\hfill September 1997\\
\end{titlepage}
\vfill
\eject

\newpage

\section{Introduction}

A large class of supersymmetric soliton solutions in string  theory
have by now been constructed in various dimensions
(for a review see for instance \cite{DuRaLu,russo,Gaunt}
and references therein), as these
play a fundamental role in duality studies. While most of the
recent work on supersymmetric solutions in string theory
has been done in dimensions higher or equal to four,
some heterotic 
supersymmetric solutions have now been determined in three space--time
dimensions \cite{sen1,BBC}.  The solutions presented in 
\cite{sen1,BBC} are static
supersymmetric solutions of the low-energy effective field theory
of heterotic string theory compactified on a seven-torus, which
is described by a three dimensional
supergravity theory 
with eight local supersymmetries \cite{marcus,sen1}.

A particular class of such heterotic solutions in three dimensions can
be obtained \cite{sen1} by compactifying  the four-dimensional 
string solutions of \cite{DGHR,DFKR} 
on a circle.  In \cite{sen1} it was shown 
that the resulting three dimensional solutions can be turned into finite
energy solutions by utilizing a mechanism first discussed in the context
of four-dimensional stringy cosmic string solutions \cite{GSVY}.  It was
further conjectured in \cite{sen1} 
that this mechanism should also apply to other 
three-dimensional supersymmetric solutions.  We will see that the same
mechanism can indeed be used for turning the solutions constructed
in \cite{BBC} into finite energy solutions.

The construction of the supersymmetric solutions given in \cite{BBC}
was achieved by
solving the associated Killing spinor equations in three dimensions.
These Killing spinors have a priori 16 real degrees of freedom which, 
however, get reduced by imposing certain contraints specific to each of
the solutions.  Up to three such independent conditions 
($m=1,2,3$) can be imposed 
on the Killing spinors, resulting in Killing spinors with $1/2^m$ of 16
real degrees of freedom.  The associated solutions were referred to 
as preserving $1/2^m$ of $N=8, D=3$ supersymmetry.

The solutions constructed in \cite{BBC} are, however, only valid 
asymptotically, that is at large spatial distances. 
Thus, they 
should get modified in such a way as to render them well behaved 
at finite distances \cite{sen1}. We will see that this is indeed 
possible by turning them into
finite energy solutions.

The solutions constructed in \cite{sen1,BBC} have a ten-dimensional
heterotic interpretation in terms of intersections of  fundamental
strings, NS $5$-branes, waves and Kaluza--Klein monopoles (or,
equivalently, an eleven dimensional interpretation in terms of
intersections of 
M-branes, M-waves and M-monopoles).  Let us for instance consider 
the solutions constructed in \cite{BBC}.  They fall into two classes.
Namely, they either carry one or two electric charges.
We will show below that the solutions carrying one electric charge
have a ten-dimensional interpretation in terms of 
 a wave and up to three orthogonally intersecting NS $5$-branes.  
The solutions
carrying two electric charges, on the other hand, have a ten-dimensional
interpretation in terms of 
a fundamental string, a wave and up to
three orthogonally intersecting NS $5$-branes as well as up to 
three Kaluza--Klein monopoles.
On the other hand, the three-dimensional solutions obtained \cite{sen1}
by compactifying  the four-dimensional heterotic string
solutions of \cite{DGHR,DFKR} on a circle have a ten-dimensional
interpretation in terms of 
a fundamental string and up to three orthogonally intersecting 
NS $5$-branes. We note here that the solutions constructed in
\cite{sen1,BBC} should also have an equivalent description in terms of
configurations of type IIB p-branes (such as the 7-brane of \cite{ggp})
wrapped around $K3\times T_3.$

The ten dimensional 
space--time line element describing 
orthogonally intersecting strings, $5$-branes, waves and Kaluza--Klein
monopoles are given in terms of harmonic functions
which  depend on some of the overall transverse
directions.  These line elements, when
compactified down to three dimensions, give rise to three dimensional
space--time line elements which are again given in terms of the same
harmonic functions.  In three space--time dimensions, a harmonic
function $H(z,\bar{z})$ does not however get determined by the 
condition of it being
harmonic.  Denoting the two spatial dimensions by $z$ and $\bar z$, one has
\beqa
\partial_z \partial_{\bar z} H = 0 \rightarrow H = f(z) + {\bar f} ( {\bar
  z}) \;\;\;,
\eeqa
where $f(z)$ is an a priori arbitrary holomorphic function.  Thus, we
expect that those heterotic supersymmetric three-dimensional solutions
which have an M-theory description in terms of orthogonally intersecting
M-branes, M-monopoles and M-waves should also be expressed in terms 
of holomorphic functions $f(z)$.
Demanding that these supersymmetric solutions  have a certain behaviour
at spatial infinity ($z \rightarrow \infty$) determines the form of $f(z)$
at large $z$.  For instance, the solutions presented in \cite{BBC} have an
asymptotic behaviour corresponding to $f(z) \propto \ln z$.  At finite
distance these asymptotic solutions  become
ill-defined
and so  need to be modified.  The associated corrections will all be
encoded in $f(z)$.  Requiring the 
solutions to have finite energy as well as the above asymptotic behaviour
will determine $f(z)$ to be given by 
$f(z) \propto j^{-1}(z)$.  The 
resulting modified solutions are then well-behaved at
finite
distances.  Asymptotically, the associated coupling constant $g^{2} 
=e^{2 \phi}$ is weak, whereas at finite distances it becomes strong.

This paper is organised as follows.  
In section 2 we review some properties 
of the low-energy effective action of heterotic string theory
compactified on a seven-dimensional torus \cite{sen1}.
In section 3 we give the Killing spinor equations associated
to the three-dimensional heterotic low-energy effective Lagrangian
and we present some results.

In section 4 we first review some of the static supersymmetric solutions
carrying two electric charges, whose asymptotic form was given in 
\cite{BBC}.  
These solutions are labelled by an integer $n$ with
$n=1,2,3,4$.
We then show how these solutions can be rewritten in terms of
holomorphic functions $f(z)$, and  check that the associated
three-dimensional Killing spinor equations are solved for any arbitrary
holomorphic $f(z)$.  Demanding that these solutions  have finite energy as 
well as the asymptotic behaviour that was found in \cite{BBC} determines 
${\hat f}(z) = {n+1 \over \pi} f(z)$
to be given by $j({\hat f}(z)) = z$.  The dilaton is, for any $n$,
determined
to be 
\beqa
e^{- \phi} = f(z) + {\bar f} ({\bar z}) \;\;\;.
\eeqa
Next, we compute the energy $E$ carried by these solutions and  find that
(in units where $8 \pi G_N =1$) $E = 2n \,{\pi \over 6}$.
The solutions presented in \cite{BBC} describe one-center solutions.  They
can be straightforwardly generalized to multi-center solutions via
$j({\hat f}(z)) = P(z)/Q(z)$ \cite{GSVY}, 
where $P(z)$ and $Q(z)$ are polynomials in
$z$ with no common factors.

In section 5, 
we repeat the analysis given in section 4 for
some of the static solutions now
carrying one electric charge with asymptotic form given in 
\cite{BBC}.  
These solutions are again labelled by an integer $n$ with
$n=1,2,3,4$.  As in the case of two electric charges, we
show how these solutions can be rewritten in terms of
holomorphic functions $f(z)$, and we check again that the associated
Killing spinor equations are solved for any arbitrary
holomorphic $f(z)$.  As before, 
demanding that these solutions  have finite energy as 
well as the asymptotic behaviour found in \cite{BBC} determines 
${\hat f} (z) = {n+2 \over 2 \pi} f(z)$
to be again given by $j({\hat f}(z)) = z$.  This time, however,
the dilaton is, for any $n$,
determined
to be
\beqa
e^{- 2\phi} = f(z) + {\bar f} ({\bar z}) \;\;\;.
\eeqa
We find that the energy $E$ associated with these solutions is (in  
units where $8 \pi G_N=1$) 
$E = n \, {\pi \over 6}$ .
This is half of the amount carried by the solutions with two electric
charges.
These solutions 
can again be straightforwardly generalized to multi-center solutions via
$j({\hat f}(z)) = P(z)/Q(z)$.

In section 6, we present the eleven-dimensional interpretation
in terms of orthogonally intersecting M-branes, M-waves and 
M-monopoles for the solutions
discussed in sections 4 and 5.

Finally, in section 7, we present our conclusions. 

We use the same conventions as in \cite{BBC}.

\section{The three-dimensional effective action}

\setcounter{equation}{0}

The effective low-energy field theory of the ten-dimensional heterotic string
compactified on a seven-dimensional torus is obtained from reducing the
ten-dimensional $N=1$ supergravity theory coupled to $U(1)^{16}$ super 
Yang--Mills multiplets 
 (at a generic point in the moduli space) \cite{fer,mah,sen1}.
The massless ten-dimensional bosonic fields are the metric $G^{(10)}_{MN}$, 
the antisymmetric tensor field $B^{(10)}_{MN}$, the $U(1)$ gauge 
fields $A^{(10)I}_M$
and the scalar dilaton $\Phi^{(10)}$ with $(0\leq M,N\leq 9,\quad 
 1\leq I\leq 16)$.
The field strengths are $F^{(10)I}_{MN}=\del_MA^{(10)I}_N-\del_NA^{(10)I}_M$ 
and $H^{(10)}_{MNP}=(\del_MB^{(10)}_{NP}-\half A^{(10)I}_MF^{(10)I}_{NP})+$
cyclic permutations of $M,N,P$.

The bosonic part of the ten-dimensional action is
\eqn
{\cal S} &\propto & \int d^{10}x\sqrt{-G^{(10)}}
e^{-\Phi^{(10)}}[{\cal R}^{(10)}
+G^{(10)MN}\del_M\Phi^{(10)}\del_N\Phi^{(10)}\no\\
&& \qquad\qquad\qquad -{1\over 12}H^{(10)}_{MNP}H^{(10)MNP}
-\hhalf F^{(10)I}_{MN}F^{(10)IMN}].
\enq
The reduction to three dimensions \cite{marcus,mah,sen1} introduces 
the graviton
$g_{\mu\nu}$, the dilaton $\phi \equiv \Phi^{(10)} -\ln \sqrt{\det G_{mn}}\,$,
 with $G_{mn}$
the internal 7D metric, 30 $U(1)$ gauge fields 
$A^{(a)}_\mu\equiv (\,A_\mu^{(1)m},\, A^{(2)}_{\mu m},\, A_\mu^{(3)I}\,)
\quad (a=1,\dots,30,\;m=1,\dots ,7,\; I=1,\dots ,16)\;,$
where $\;A_\mu^{(1)m}\;$ are the 7 Kaluza--Klein gauge fields coming from 
the reduction of $G_{MN}^{(10)},\;A_{\mu m}^{(2)}\equiv B_{\mu m} 
+ B_{mn}A_\mu^{(1)n}+\half a_m^IA_\mu^{(3)I}\;$ are the 7 gauge fields coming
from the reduction of $\;B_{MN}^{(10)}\;$ and $\;A_\mu^{(3)I}\equiv A_\mu^I\
-a_m^IA_\mu^{(1)m}\;$ are the 16 gauge fields from  $A_M^{(10)I}$.

 The field strengths $F_{\mu\nu}^{(a)}$ are given by 
$F_{\mu\nu}^{(a)}=\del_\mu A_\nu^{(a)}-\del_\nu A_\mu^{(a)}$. 
Finally, $\;B_{MN}^{(10)}\;$ induces the two-form field 
$\;B_{\mu\nu}\;$ with field strength $\;H_{\mu\nu\rho}=\del_\mu 
B_{\nu\rho}-\half A_\mu^{(a)}L_{ab}F_{\nu\rho}^{(b)}$+ cyclic permutations.
 
 The 161 scalars $G_{mn},\,a_m^I$ and $B_{mn}$ can be arranged 
into a $30\times 30$ matrix $M$ (we use here the conventions of \cite{mah})
\eq
M=  \left( \begin{array}{ccc}
G^{-1} & -G^{-1}C & -G^{-1}a^T \\
-C^TG^{-1} & G+C^TG^{-1}C+a^Ta & C^TG^{-1}a^T+a^T\\
-aG^{-1} & aG^{-1}C+a & I_{16}+aG^{-1}a^T
\end{array} \right)\;\;,\label{M} 
\en
where $G\equiv [{G}_{mn}],\,\, C\equiv [\half 
a_m^{I}a_n^{I}+B_{mn}]$ and $a\equiv [a^I_m]$.

We have $MLM^T=L,\quad M^T=M,\quad L^{-1}=L,$ where
\eq
 L =  \left( \begin{array}{ccc}
0 & I_7 & 0\\
I_7 & 0 & 0\\
0 & 0 & I_{16} 
\end{array} \right)\;\;. 
\en
    We use the following ansatz for the Kaluza--Klein 10D vielbein 
$E_M^A$ and inverse vielbein $E_A^M$, in the string frame
\eq\label{viel}
\ E^A_M=  \left( \begin{array}{cl}
e^{\phi}e^\alpha_\mu & A_\mu^{(1)m}e^a_m \\
0 & e^a_m 
\end{array} \right)\;\;,\qquad E_A^M=  \left( \begin{array}{cl}
e^{-\phi}e_\alpha^\mu & -e^{-\phi}e_\alpha^\mu A_\mu^{(1)m} \\
0 & e_a^m 
\end{array} \right)\;\;, 
\en
where $e_m^a$ is the internal  and $e_\mu^\alpha$  
the space--time vielbein in the Einstein frame (the relation between
 string metric $G_{\mu\nu}$ and  Einstein metric $g_{\mu\nu}$
in three dimensions is $G_{\mu\nu}=e^{2\phi}g_{\mu\nu}$).

The three-dimensional action in the Einstein frame is then \cite{mah,sen1}, 
\eqn
{\cal S}&=&\hhalf\int d^3x\sqrt{-g}\bigl\{{\cal R}-g^{\mu\nu}
\del_\mu\phi\del_\nu\phi - 
{1\over 12} e^{-4\phi}g^{\mu\mu'}g^{\nu\nu'}g^{\rho\rho'}H_{\mu\nu\rho}
H_{\mu'\nu'\rho'}\no\\
&& - \hhalf e^{-2\phi}g^{\mu\mu'}g^{\nu\nu'}F_{\mu\nu}^{(a)}
(LML)_{ab}F_{\mu'\nu'}^{(b)}+ {1\over 8}g^{\mu\nu}\Tr (\del_\mu ML\del_\nu
ML)\big\}\;\;,
\label{act3d}
\enq
where $a=1,\dots,30.$

This action is invariant under the $O(7,23)$ transformations
\eq
M\rightarrow \tilde{\Omega}M\tilde{\Omega}^T,\quad A_\mu^{(a)}\rightarrow
\tilde{\Omega}_{ab}A_\mu^{(b)},\quad g_{\mu\nu}\rightarrow g_{\mu\nu},\quad
B_{\mu\nu}\rightarrow B_{\mu\nu},\quad \phi\rightarrow\phi,\quad
\tilde{\Omega}^TL\tilde{\Omega}=L,\label{07}
\en
where $\tilde{\Omega}$ is a $30\times 30$ $O(7,23)$ matrix.

The equations of motion for $A^{(a)}_\mu,\,\, \phi,$ $H^{\mu\nu\rho}$ 
and $g^{\mu\nu}$ are, respectively,
\eqn
&&\del_\mu(e^{-2\phi}\sqrt{-g}(LML)_{ab}
F^{(b)\mu\nu})+\half e^{-4\phi}\sqrt{-g}\,L_{ab}\,F^{(b)}_{\mu\rho}
H^{\nu\mu\rho}=0\;,\label{ff} \label{eqfmn}\\
&&D_\mu D^\mu\phi +\hhalf e^{-2\phi}F_{\mu\nu}^{(a)}(LML)_{ab}
F^{\mu\nu (b)}+{1\over 6}e^{-4\phi}H^{\mu\nu\rho}H_{\mu\nu\rho}=0\;,
\label{phi}\\
&&\del_\mu(\sqrt{-g}e^{-4\phi}H^{\mu\nu\rho})=0\;,\label{H}\\
&&{\cal R}_{\mu\nu}=\del_\mu\phi\del_\nu\phi + \half e^{-2\phi}
F_{\mu\rho}^{(a)}(LML)_{ab}
F_\nu^{\rho (b)}-{1\over 8}\Tr (\del_\mu ML\del_\nu
ML)\label{Ricci}\label{eqrmn} \\
&&\qquad-\hhalf e^{-2\phi}g_{\mu\nu}F_{\rho\tau}^{(a)}(LML)_{ab}
F^{\rho\tau (b)}+\hhalf e^{-4\phi}H_\mu^{\tau\sigma}H_{\nu\tau\sigma}
-{1\over 6}g_{\mu\nu}e^{-4\phi}H^{\tau\sigma\rho}H_{\tau\sigma\rho}
\;.\no
\enq
We will, in the following, consider backgrounds where
$H_{\mu\nu\rho}=0$. 

Consider a static solution with space--time line element of the 
form
\beqa
ds^2 = -dt^2 + g_{xx}\,(dx^2 + dy^2) = - dt^2 + 2 g_{z{\bar z}}\, 
dz d{\bar z} \;\;\;,\;\;\; 
\eeqa
where $z= x+iy$.  Its energy $E$ can be computed from  
 (\ref{eqfmn})--(\ref{eqrmn}) and is given by
\beqa
E &=&{1\over{8\pi G_N}}\int d^2x  \,g_{xx} T_{tt}
={i\over{8\pi G_N}}\int dz d{\bar z}  \, g_{z{\bar z}} T_{tt}
= {i\over{16\pi G_N}} \int dz d{\bar z}
\,g_{z{\bar z}}{\cal R}\;\;, \label{ener}
\eeqa
where $8\pi G_N \,T_{\mu\nu} = {\cal R}_{\mu\nu} -\half \,g_{\mu\nu}{\cal R}$.
Here $G_N$ denotes the three-dimensional gravitational constant.  We will
in the following set $8\pi G_N =1$.

{From} the equations of motion for the gauge
fields $A_\mu^{(a)}$ (\ref{ff}) one can define a  set of scalar 
fields $\Psi^a,\; a=1,\dots,30 ,$ through \cite{sen1}
\eqn
&&\sqrt{-g}e^{-2\phi}g^{\mu\mu'}g^{\nu\nu'}(ML)_{ab}F_{\mu'\nu'}^{(b)}=
 \epsilon^{\mu\nu\rho}\del_\rho\Psi^a,\no\\
&&F^{(a)\mu\nu}={1\over{\sqrt{-g}}}e^{2\phi}(ML)_{ab}
\epsilon^{\mu\nu\rho}\del_\rho\Psi^b\label{psi}\;.\label{sca}
\enq
Then, from the Bianchi identity $\epsilon^{\mu\nu\rho}\del_\mu
F_{\nu\rho}^{(a)}=0$,
\eq
D^\mu(e^{2\phi}(ML)_{ab}\del_\mu\Psi^b)=0.
\en
Following \cite{sen1},
the charge quantum numbers of elementary string
excitations are characterized by a 30-dimensional vector $\vec{\alpha}\in 
\Lambda_{30}$. The asymptotic value of the field strength $F_{\mu\nu}^{(a)}$
associated with such an elementary particle can be calculated to be \cite{sen1}
\eq
\sqrt{-g}F^{(a)tr}\simeq -{1\over{2\pi}}e^{2\phi}M_{ab}\alpha^b.\label{F0r}
\en
The asymptotic form of $\Psi^a$ is then
\eq
\Psi^a\simeq -{\theta\over {2\pi}}L_{ab}\alpha^b + {\rm constant}.
\label{asympsi}
\en
It can be shown \cite{sen1} that the matrix 
$M$, the $\Psi$'s and the dilaton can
be assembled into a matrix ${\cal M}$ describing the coset
$\frac{O(8,24)}{O(8)  \times O(24)}$.  The
low energy effective three-dimensional 
field theory is then actually invariant under $O(8,24)$ transformations.
An $O(8,24;\IZ)$ subgroup of this group is 
a symmetry of the full string theory \cite{sen1}.

\section{The Killing spinor equations}

\setcounter{equation}{0}

In ten dimensions, the supersymmetry transformation rules  for 
the gaugini $\chi^I$, dilatino
$\lambda$ and gravitino $\psi_M$ are,  in the string frame, given by 
\cite{julia, stro, cand, liu, peet}
\eqn
&&\delta \chi^I= \half F^I_{MN}\Gamma^{MN}\varepsilon\;\;,\no\\
&&\delta\lambda =-\half \Gamma^M\del_M\Phi\varepsilon
 +{1\over 12} H_{MNP}\Gamma^{MNP}\varepsilon\;\;,\no\\
&&\delta\psi_M=\del_M\varepsilon + {1\over 4}(\omega_{MAB}-\half H_{MAB})
\Gamma^{AB}\varepsilon\;\;.
\enq
These equations were reduced to three dimensions in the Einstein
frame in \cite{BBC}.  In the following, we  restrict ourselves to
backgrounds with $H_{\mu\nu\rho}= 0$ and $a_m^I=0$.  The associated
three-dimensional Killing spinor equations become 
\eqn
\delta \chi^I&=&\half e^{-2\phi} F^{(3)I}_{\mu\nu}
\gamma^{\mu\nu}\varepsilon ,\no\\
\delta\lambda &=& -\half e^{-\phi}\del_\mu \{\phi+\ln\det e_m^a\} 
\gamma^\mu\otimes{\bf I}_8\,\varepsilon 
+\hhalf e^{-2\phi}[-B_{mn}F^{(1)n}_{\mu\nu}+F^{(2)}_{\mu\nu m} 
]\gamma^{\mu\nu}\gamma^4\otimes\Sigma^m\varepsilon\no\\
&&+\hhalf e^{-\phi}\del_\mu B_{mn}\gamma^\mu\otimes\Sigma^{mn}
\varepsilon\;\;,\no\\
\delta{\psi}_\mu &=&\del_\mu\varepsilon + {1\over 4}\omega_{\mu\alpha\beta}
\gamma^{\alpha\beta}\varepsilon 
 +\hhalf (e_{\mu\alpha}e_\beta^\nu \!-\!e_{\mu\beta}e_\alpha^\nu)
\del_\nu\phi\gamma^{\alpha\beta}\varepsilon + {1\over 8}(e^n_a\del_\mu 
e_{nb}\!-\!e^n_b\del_\mu e_{na}){\bf I}_4\otimes\Sigma^{ab}\varepsilon \no\\
&&
\!-\!{1\over 4}e^{-\phi}[e^m_a F^{(2)}_{\mu\nu(m)} 
- e_{ma} F^{(1)m}_{\mu\nu}]\gamma^\nu\gamma^4\otimes\Sigma^a\varepsilon
\!-\!{1\over 8}\del_\mu B_{mn}\! \;{\bf I}_4
\otimes\Sigma^{mn}\varepsilon \no\\
&&+\hhalf e^{-\phi}B_{mn}F_{\mu\nu}^{(1)n}\;\gamma^\nu\gamma^4\otimes
\Sigma^m\varepsilon\;\;,\no\\
\delta\psi_d &=&-{1\over 4}e^{-\phi}(e_d^m\del_\mu e_{ma}\!+\!e_a^m\del_\mu 
e_{md})\gamma^\mu\gamma^4\otimes\Sigma^a\varepsilon
+{1\over 8}e^{-2\phi}e^m_d B_{mn}F^{(1)n}_{\mu\nu} \;
\gamma^{\mu\nu}\varepsilon\no\\
&&+{1\over 4}e^{-\phi}e^m_de^n_a \del_\mu B_{mn}\gamma^\mu
\gamma^4\otimes\Sigma^a\varepsilon
-{1\over 8}e^{-2\phi}[\;e_{md}F_{\mu\nu}^{(1)m}
+e_d^mF^{(2)}_{\mu\nu m}\;]\gamma^{\mu\nu}\varepsilon\;\;,\label{killing}
\enq
where $\delta\psi_d\equiv e_d^m\delta\psi_m$ denotes the variation of the
internal gravitini.

Static solutions to the Killing spinor 
equations can be constructed
by setting the supersymmetry variations of the fermionic fields 
to zero. This ensures that the bosonic configurations so obtained 
are supersymmetric.

The supersymmetric static solutions 
we will be discussing in the following sections
will either carry one or two electric charges.  They
have the following space--time line elements:
\eq
ds^2=- dt^2 + H^{cn}
 (\omega, {\bar \omega}) d\omega d{\bar \omega}\;\;\;,\;\;\; c=1,2
 \;\;\;,\;\;\;
n=1,2,3,4 \;\;\;,
\label{gg}
\en
where $\omega = \hat{r} + i \hat{\theta}$, and where 
the solutions carrying one (two) electric charge
have $c=1$ ($c=2$).  
$H(\omega, {\bar \omega})$ 
denotes a harmonic function:
\beqa
\partial_{\omega} \partial_{\bar \omega} H = 0 \longrightarrow H(\omega
, {\bar \omega}) = f(\omega) + f({\bar \omega}) \;\;\;.
\eeqa
The dilaton is found to be
\beqa
e^{- 2 \phi} = H^c \;\;\;.
\eeqa

In all cases, we will make the following ansatz for the Killing spinors
\eq
\varepsilon=\epsilon\otimes\chi\label{spinor}\;\;,
\en
where $\epsilon^T=(\epsilon_1,\,\epsilon_2,\,\epsilon_3,\,\epsilon_4)$ is a
$SO(1,2)$ spinor and $\chi$ is a $SO(7)$ spinor of the internal space. 
We will be able to solve the 
Killing spinor equations 
by imposing the following two conditions on $\epsilon$ \cite{BBC}:
\eq\label{cond1}
\gamma^1\epsilon =ip\;\gamma^2\;\epsilon\;\;,\qquad\qquad
\gamma^1\epsilon =\tilde{p}\;J\gamma^2\gamma^4\;\epsilon\;\;,
\en
where $p=\pm\;,\;\;\tilde{p}=\pm\;.$  Here we used that $\gamma^{\mu}
= \gamma^{\alpha} e_{\alpha}^{\mu} \,,\, \alpha=0,1,2$.

It follows that 
\beqa
\epsilon = \tilde{\epsilon}(\omega, {\bar \omega}) \,
\left( \begin{array}{c}
 i p \\
1 \\
 {\tilde p}\\
-i p {\tilde p}
\end{array} \right) \;\;\;,\label{spinor1}
\eeqa
and, hence, $\epsilon$ contains only two real independent degrees of
freedom. $\chi$, on the other hand, contains eight real degrees of freedom;
thus there are a priori a total of 16 real degrees of freedom. These will
be further reduced by conditions on $\chi$ specific to each case
considered. Up to three such independent
conditions $(m=1,2,3)$ can be imposed on
$\chi$, thereby allowing for the construction of solutions whose Killing
spinors have $1/{2^m}$ of 16 real degrees of freedom.  The 
solutions with $n=1$ and $n=2$ have Killing spinors with $8$ and $4$ real
degrees of freedom, respectively.  The solutions with $n=3$ and $n=4$ both have
Killing spinors with $2$ real degrees of freedom.

In all cases considered in the following, we find that
\eq
\tilde{\epsilon}=e^{{\phi}\over 2} = H^{- \frac{c}{4}}\;\;,\label{spinor2}
\en
up to a multiplicative constant.

\section{Supersymmetric solutions carrying two electric charges }\label{two}

\setcounter{equation}{0}

Here, we will consider the class of solutions 
carrying two
electric charges 
presented in \cite{BBC}.  The solutions given there are
well behaved only at large
spatial distances,
and hence need to be
modified at finite distances.  

The solutions in this class are labelled by an integer $n$ ($n=1,2,3,4$).
The associated
dilaton was given by (in a specific coordinate system) \cite{BBC}
\beqa
e^{ \phi} = \frac{n+1}{a \ln r}  \;\;\;,
\label{dilasy}
\eeqa
whereas the associated space--time line element 
was  
\beqa
ds^2=-dt^2+ \frac{a^2}{r^2} \; \left(\frac{ a \ln r }{n+1} \right )^{2n}
(dr^2 + r^2 d\theta^2).
\label{oldline}
\eeqa
Here, $a = \frac{n+1}{2 \pi} \sqrt{|\alpha_i \alpha_{i + 7}|}$, where
$\alpha_i$ and $\alpha_{i + 7}$ denote the two electric charges.
The i-th component of the internal metric $G_{mn}$ was  given 
by $G_{ii} = |\frac{\alpha_i}{\alpha_{i+7}}|$.  The gauge fields strengths,
or equivalently $\Psi^i$ and $\Psi^{i+7}$, were taken to have the
asymptotic
form given in (\ref{asympsi}).
Introducing complex coordinates,  
$\omega = a \ln z = a( \ln r + i \theta) $,
then yields
\beqa\label{cc1}
ds^2 = 
-dt^2 + H^{2n} d\omega d{\bar \omega} \;\;\;, \;\;\; e^{- \phi} = H \;\;\;,
\label{linm}
\eeqa
where $H = \frac{a \ln r}{n+1} = \frac{\omega + \bar \omega}{2(n+1)}$.
Inspection of (\ref{dilasy}) shows that the solution becomes ill defined
as $r \rightarrow 1$ and thus it should get modified at finite distances.
The asymptotic form of $H$ suggests the replacement
\beqa
H =  \frac{\omega + \bar \omega}{2(n+1)} \longrightarrow H = f(\omega) + 
{\bar f}( {\bar \omega}) \;\;\;.
\label{hmod}
\eeqa
Similarly, 
the asymptotic form of the $\Psi^i$ given in (\ref{asympsi}) suggests
the replacement
\beqa
\Psi^i = \frac{ \alpha_{i+7}}{4 \pi a } \; i (\omega - {\bar \omega})
 \longrightarrow
\Psi^i = 
\frac{(n+1) \alpha_{i+7}}{2 \pi a } \; i  (f - {\bar f}) \;\;\;,\no\\
\Psi^{i+7} = \frac{ \alpha_{i}}{4 \pi a } \; i (\omega - {\bar \omega})
 \longrightarrow
\Psi^{i+7} = 
\frac{(n+1) \alpha_{i}}{ 2 \pi a}  \; i  (f - {\bar f}) \;\;\;.
\label{newpsi}
\eeqa
Denoting the imaginary part of $f$ by $\hat \Psi$, $\hat \Psi = - i
(f - \bar f)$, we have 
\beqa
f = \frac{1}{2} \left(e^{- \phi} + i \hat \Psi
\right) \;\;\;.
\eeqa
Using  $G_{ii} = |\frac{\alpha_i}{\alpha_{i+7}}|$, we 
can rewrite
$\Psi^i$ and $\Psi^{i + 7}$ in the following way
\beqa
\Psi^i = - \eta_{\alpha_{i+7}} \sqrt{G^{ii}} {\hat \Psi} \;\;\;, \qquad \qquad
\Psi^{i+7} =  - \eta_{\alpha_i} \sqrt{G_{ii}} {\hat \Psi} \;\;\;,
\eeqa
where $\eta_{\alpha_{i+7}}= - \eta_{\alpha_i}$ denotes the sign of the 
two charges ${\alpha_i}$ and ${\alpha_{i+7}}$.
Thus we see that the $\Psi$'s can be reexpressed in terms of
the internal metric and of ${\hat \Psi}$.  

Using that $\partial_{\bar \omega} f = 0$ it can be shown that the 
associated field strengths are given by
\beqa\label{fs1}
F_{t \beta}^{(1)i} = \eta_{\alpha_i} \sqrt{G^{ii}} \;
\frac{\partial_{\beta}H}{H^2}\;\;\;,
\qquad\qquad F_{t \beta \,i}^{(2)} = - \eta_{\alpha_i} \sqrt{G_{ii}} \;
\frac{\partial_{\beta}H}{H^2} \;\;\;,\;\;\; \beta = {\hat r}, {\hat
  \theta} \;\;\;,
\label{gf2}
\eeqa
where $\omega = a(\ln r + i \theta) = {\hat r} + i {\hat \theta}$.

We  now discuss the solutions in more detail.
Let us first consider the case where $n=1$.  The associated Killing spinor has
$1/2$ of $16$ real degrees of freedom.  The solution 
is characterised by the fact that 
the internal vielbein
$e_m^a$ is diagonal and constant and that $B_{mn}=0$ \cite{BBC}.
The solution presented in \cite{BBC}, which has
line element (\ref{oldline}), should be modified so as to render the solution
well behaved at finite distances.  This modification is given by
(\ref{linm}), (\ref{hmod}) and (\ref{newpsi}).  It can then be checked that
the Killing spinor equations (\ref{killing}) are satisfied for any arbitrary
holomorphic $f$.  The Killing spinor is given by (\ref{spinor}) 
and (\ref{spinor1})
subject to one additional condition on $\chi$ given in $\cite{BBC}$.
Thus, the requirement of supersymmetry alone does not determine $f$.

Next, consider the case where $n=2$.
The associated Killing spinor has
$1/4$ of $16$ real degrees of freedom.  The solution 
is characterised by the fact that now
there is one non-constant off-diagonal entry in the internal metric
as well as one non-constant entry in the $B_{mn}$-matrix.
As an example, consider the case where 
the two electric charges are taken to be 
$\alpha_4$ and $\alpha_{11}$, and where the background fields $G_{mn}$ and
$B_{mn}$ are given by \cite{BBC}
\renewcommand{\arraystretch}{0.6}
\beqa
\label{gbdyon}
\left(\!\!\! \begin{array}{cccccc}
{G}^{11} & {G}^{12}& 0 & 0 &  \cdots & 0\\
{G}^{21} & {G}^{22}& 0 & 0 & \cdots & 0 \\
0 & 0 & {G}^{33} & 0 & \cdots    &  \\
0 & 0 & 0 &  {G}^{44} & 0  &  \vdots \\ 
\vdots &  &   &  & \ddots &   \\
0& & \cdots &  & 0 & {G}^{77}
\end{array}\!\!\! \right)
\!\!=\!\!
\left(\!\!\!\! \begin{array}{cccccc}
g^2  & - g^2 \Upsilon_2  & 0 &0& \cdots & 0\\
- g^2 \Upsilon_2 & |\frac{\alpha_9}{\alpha_2}| 
+  g^2 \Upsilon_2^2 & 0& 
0 & \cdots & 0 \\
0 & 0 & {G}^{33} & 0 &  \cdots &  \\
0 & 0 & 0 &  {G}^{44} & 0  & \vdots  \\ 
\vdots & & & &  \ddots  \\
0& & \cdots  &  & 0& {G}^{77}
\end{array}\!\!\! \right)\!, &&\nonumber\\
\no\\
\renewcommand{\arraystretch}{1.0}
{ B} =  ({ B}_{mn}) = 
\left( \begin{array}{ccccc}
0 &{B}_{12} & 0 & \cdots &0\\
{ B}_{21} & 0 \\
0& & 0 & & \vdots\\
\vdots& & & \ddots \\
0& &\cdots & & 0
\end{array} \right)\!\!=\!\! 
\left( \begin{array}{ccccc}
0 &  \Upsilon_9 & 0 & \cdots &0\\
- \Upsilon_9 & 0 \\
0& & 0 & & \vdots\\
\vdots& & & \ddots \\
0& &\cdots & & 0
\end{array} \right) ,\qquad\qquad\qquad &&
\eeqa
where $G^{44} = |\frac{\alpha_{11}}{\alpha_4}| $.
The solution presented in \cite{BBC} is valid only at large distances, and
again it
should be modified at finite
distances.  These modifications are given by 
(\ref{linm}), (\ref{hmod}) and (\ref{newpsi}) as well as by 
\renewcommand{\arraystretch}{1.0}
\beqa
g = \frac{D}{H}  \;\;\;,\;\;\;
\Upsilon_2 = 
\frac{(n+1) \alpha_9}{2 \pi a } \; i  (f - {\bar f}) \;\;\;,\;\;\;
\Upsilon_9 = 
\frac{(n+1) \alpha_2}{2 \pi a } \; i  (f - {\bar f}) \;\;\;,
\eeqa
where $D = \frac{\sqrt{|\alpha_4 \alpha_{11}|}}{
\sqrt{|\alpha_2 \alpha_{9}|}}$.  It can again be checked that this modified
background satisfies the Killing spinor equations (\ref{killing}) for
any arbitrary holomorphic $f$.   The associated Killing spinor is given
by (\ref{spinor}) and (\ref{spinor1}) subject to two additional constraints
on $\chi$ given in \cite{BBC}.

Finally, consider the cases where $n=3$ and $n=4$.  The associated Killing
spinors both have
$1/8$ of $16$ real degrees of freedom.  The solutions 
are characterised by the addition of one (two) additional off-diagonal
entries in the internal metric $G_{mn}$ and in $B_{mn}$ for $n=3$ ($n=4$)
\cite{BBC}.
It is straightforward to modify these solutions along similar lines
as the ones discussed above, and again it can be checked that these
modified solutions satisfy the Killing spinor equations (\ref{killing}).

We thus see that, in any of the above modified solutions,
the modifications are all encoded in one single holomorphic function $f$.

Next, we would like to determine the holomorphic function $f$ by demanding
that the modified solution  have finite energy \cite{sen1}.  
This will also render the solutions well behaved at finite distances.

Let us first compute the energy carried by the modified 
solutions discussed above.  The integral (\ref{ener}) is computed to be
(in units where $8 \pi G_N =1$)
\beqa
E &=&i \, 2n 
\int d\omega d{\bar \omega}\, \frac{\partial_{\omega}f \partial_{\bar \omega}
{\bar f}}
{(f + {\bar f})^2}
= i \, 2n \int dz d{\bar z}\, 
\frac{\partial_{z} {\hat f} 
\partial_{\bar z}{\bar {\hat f}}}
{({\hat f} + {\bar {\hat f}})^2} \;\;\;,
\label{enint}
\eeqa
where $\omega = a\ln z$ and where we have introduced ${\hat f} =
\frac{n+1}{\pi}f$ for later convenience.  

There is an elegant mechanism \cite{GSVY}
for rendering the integral (\ref{enint})
finite.  Let us take $z$ to be the coordinate of a complex plane.  Then
there is a one-to-one map from a certain domain F on the ${\hat f}$-plane
(the so called `fundamental' domain) to the $z$-plane.  This map is known
as the j-function, $j({\hat f}) = z$.  By means of this map, the integral
(\ref{enint}) can be pulled back from the $z$-plane to the domain F
(the $z$-plane covers F exactly once).  Then, by using integration by
parts,
this integral can be related to a line integral over the boundary of F,
which is evaluated to be \cite{GSVY} 
\beqa
E = 2n\,\frac{2 \pi}{12} = 2n \,\frac{\pi}{6}
\label{ener2}
\eeqa
and, hence, is finite.

By expanding $j({\hat f}) = 
e^{2 \pi {\hat f}} + 744 + {\cal O} ( e^{-2 \pi {\hat f}} ) = z$ we see that
$f(\omega) = \frac{1}{2(n+1)} \left( \omega - 744 e^{- \omega} + \dots
 \right) $, which indeed reproduces the correct asymptotic behaviour at
$\omega \rightarrow \infty$.  Thus we see that,  by demanding 
 the asymptotic
behaviour of $f$ to be modified to $f = \frac{\pi}{n+1} \, j^{-1}(z)$,
the associated energy becomes finite.

We note that the solutions discussed above represent one-center
solutions.  They can be generalised to multi-center solutions via
$j({\hat f}(z)) = P(z)/Q(z)$, where $P(z)$ and $Q(z)$ are polynomials in
$z$ with no common factors.  These are the analogue of the multi-string
configurations discussed in \cite{GSVY}.

It can be checked that the curvature scalar ${\cal R} \propto
\partial_{\omega} f \partial_{\bar \omega} {\bar f}$ 
blows up at the special point
${\hat f}=1$ (at this point, the $j$-function and its derivatives are given
by $j = 1728, j' =0$), 
whereas it is well behaved at the point 
${\hat f} = 
e^{i \pi/6}$  (at this point, the $j$-function and its derivatives are given
by $j  = j'= j''=0$).

\section{Supersymmetric solutions carrying one electric charge}

\setcounter{equation}{0}

Here we will first review the supersymmetric solutions constructed in
\cite{BBC}
carrying one electric charge.  These solutions need again to be
modified at finite distances.  

The solutions in this class are also labelled by an integer $n$
($n=1,2,3,4$).  Here the dilaton was \cite{BBC}
\beqa
e^{2 \phi} = \frac{n+2}{2 a \ln r}  \;\;\;,
\label{dilasy1}
\eeqa
whereas the associated space--time line element was 
\beqa
ds^2=-dt^2+ \frac{a^2}{r^2} \; \left(\frac{2 a \ln r }{n+2} \right )^{n}
(dr^2 + r^2 d\theta^2).
\label{oldline1}
\eeqa
Here, $a = \frac{n+2}{4 \pi} |\alpha_i|$, where
$\alpha_i$ denotes the electric charge carried by the solutions.
The i-th component of the internal metric $G_{mn}$ was  given 
by $G_{ii} =\frac{2 a \ln r }{n+2}$.  
The gauge fields strength,
or equivalently $\Psi^{i+7}$, was taken to have the
asymptotic
form given in (\ref{asympsi}).
Introducing complex coordinates,  
$\omega = a \ln z = a( \ln r + i \theta) $,
then yields
\beqa\label{cc}
ds^2 = 
-dt^2 + H^{n} d\omega d{\bar \omega} \;\;\;, \;\;\; e^{- 2\phi} = H \;\;\;,
\;\;\; G_{ii} = H \;\;\;,
\label{linm1}
\eeqa
where $H = \frac{2a \ln r}{n+2} = \frac{\omega + \bar \omega}{n+2}$.
The solution again becomes ill defined
as $r \rightarrow 1$ and thus needs to get modified at finite distances.

The asymptotic form of $H$ suggests the replacement
\beqa
H =  \frac{\omega + \bar \omega}{n+2} \longrightarrow H = f(\omega) + 
{\bar f}( {\bar \omega}) \;\;\;.
\label{hmod1}
\eeqa
Similarly, 
the asymptotic form of $\Psi^{i+7}$ given in (\ref{asympsi}) suggests
the replacement
\beqa
\Psi^{i+7} = \frac{ \alpha_{i}}{4 \pi a } \; i (\omega - {\bar \omega})
 \longrightarrow
\Psi^{i+7} = 
\frac{(n+2) \alpha_{i}}{ 4 \pi a}  \; i  (f - {\bar f}) 
= \eta_{{\alpha}_i} \, i  (f - {\bar f}) 
\;\;\;,
\label{newpsi1}
\eeqa
where 
$\eta_{\alpha_i}$ denotes the sign of the 
charge $\alpha_i$.
Denoting the imaginary part of $f$ by $\hat \Psi$, $\hat \Psi = - i
(f - \bar f)$, yields 
\beqa
f = \frac{1}{2} \left(e^{- 2\phi} + i \hat \Psi
\right) 
\eeqa
as well as 
\beqa
\Psi^{i+7} =  - \eta_{\alpha_i}{\hat \Psi} \;\;\;.
\eeqa
The 
associated field strength is given by
\beqa\label{fs}
F_{t \beta}^{(1)i} &=& \eta_{\alpha_i}  \;
\frac{\partial_{\beta}H}{H^2} \;\;\;,\;\;\; \beta= {\hat r}, {\hat
  \theta} \;\;\;,
\label{gf1}
\eeqa
where $\omega = a(\ln r + i \theta) = {\hat r} + i {\hat \theta}$.

We will now discuss the solutions in more detail.
Let us first consider the case where $n=1$.  The associated Killing spinor has
$1/2$ of $16$ real degrees of freedom.  The solution
is characterised by the fact that 
the internal vielbein
$e_m^a$ is diagonal and that $B_{mn}=0$ \cite{BBC}.
The solution presented in \cite{BBC}, which has
line element (\ref{oldline1}), should be modified to render the solution
well behaved at finite distances.  This modification is given by
(\ref{linm1}), (\ref{hmod1}) and (\ref{newpsi1}).  It can then be checked that
the Killing spinor equations (\ref{killing}) are satisfied for any arbitrary
holomorphic $f$.  The Killing spinor is given by (\ref{spinor}) 
and (\ref{spinor1}) with $p =-1$, 
subject to one additional condition on $\chi$ given in $\cite{BBC}$.
Thus, the requirement of supersymmetry alone does not determine $f$.

Next, consider the case where $n=2$.
The associated Killing spinor has
$1/4$ of $16$ real degrees of freedom.  The solution
is characterised by the fact that now
there is one non-constant entry in the $B_{mn}$-matrix.  The internal
metric $G_{mn}$ stays diagonal, however.
As an example, consider the case where 
the electric charge is taken to be 
$\alpha_4$, and where the background fields $G_{mn}$ and
$B_{mn}$ are given by \cite{BBC}
\renewcommand{\arraystretch}{0.6}
\beqa
\label{gbdyon1}
G^{-1} &=& \left(\!\!\! \begin{array}{cccccc}
{G}^{11} & 0& 0 & 0 &  \cdots & 0\\
0 & {G}^{22}& 0 & 0 & \cdots & 0 \\
0 & 0 & {G}^{33} & 0 & \cdots    &  \\
0 & 0 & 0 &  {G}^{44} & 0  &  \vdots \\ 
\vdots &  &   &  & \ddots &   \\
0& & \cdots &  & 0 & {G}^{77}
\end{array}\!\!\! \right)
\!\!\;=\!\!\;
\left(\!\!\!\! \begin{array}{cccccc}
g^2_1  & 0  & 0 &0& \cdots & 0\\
0 & g^2_2 & 0& 
0 & \cdots & 0 \\
0 & 0 & {G}^{33} & 0 &  \cdots &  \\
0 & 0 & 0 &  \frac{1}{H} & 0  & \vdots  \\ 
\vdots & & & &  \ddots  \\
0& & \cdots  &  & 0& {G}^{77}
\end{array}\!\!\! \right)\!, \nonumber\\
\renewcommand{\arraystretch}{1.0}
{ B} &=&
\left( \begin{array}{ccccc}
0 &{B}_{12} & 0 & \cdots &0\\
{ B}_{21} & 0 \\
0& & 0 & & \vdots\\
\vdots& & & \ddots \\
0& &\cdots & & 0
\end{array} \right)= 
\left( \begin{array}{ccccc}
0 &  \Upsilon_9 & 0 & \cdots &0\\
- \Upsilon_9 & 0 \\
0& & 0 & & \vdots\\
\vdots& & & \ddots \\
0& &\cdots & & 0
\end{array} \right) .\qquad\qquad\qquad 
\eeqa
The solution presented in \cite{BBC} is valid only at large distances, and
it
should be modified at finite
distances.  These modifications are given by 
(\ref{linm1}), (\ref{hmod1}) and (\ref{newpsi1}) as well as by 
\renewcommand{\arraystretch}{1.0}
\beqa
g_1^2 = \frac{D^2_1}{H}  \;\;\;,\;\;\; g_2^2 = \frac{D_2^2}{H} \;\;\;,\;\;\;
\Upsilon_9 = 
\frac{(n+2) \alpha_2}{4 \pi a } \; i  (f - {\bar f}) \;\;\;,
\eeqa
where $D_1 D_2  = \frac{4 \pi a}{(n+2) |\alpha_2|}$.
It can again be checked that this modified
background satisfies the Killing spinor equations (\ref{killing}) for
any arbitrary holomorphic $f$.   The associated Killing spinor is given
by (\ref{spinor}) and (\ref{spinor1}) with $p =-1$,
subject to two additional constraints
on $\chi$ given in \cite{BBC}.

Finally, consider the cases where $n=3$ and $n=4$.  The associated
Killing
spinors both have
$1/8$ of $16$ real degrees of freedom.  
The solutions 
are characterised by the addition of one (two) additional off-diagonal
entries in $B_{mn}$ for $n=3$ ($n=4$)
\cite{BBC}.  In both cases, the internal metric $G_{mn}$ stays diagonal.
It is straightforward to modify these solutions along similar lines
as the ones discussed above, and  it can be checked that these
modified solutions satisfy the Killing spinor equations (\ref{killing}).

The modifications are again all encoded in one single 
holomorphic function $f$.

Next, we would like to determine the holomorphic function $f$ by 
demanding
that the modified solution should have finite energy \cite{sen1}.  
This time, computing the integral (\ref{ener}) yields
(in units where $8 \pi G_N =1$)
\beqa
E &=&i \, n 
\int d\omega d{\bar \omega}\, \frac{\partial_{\omega}f \partial_{\bar \omega}
{\bar f}}
{(f + {\bar f})^2}
= i \, n \int dz d{\bar z}\, 
\frac{\partial_{z} {\hat f} 
\partial_{\bar z}{\bar {\hat f}}}
{({\hat f} + {\bar {\hat f}})^2} \;\;\;,
\label{enint1}
\eeqa
where $\omega = a\ln z$ and where this time  ${\hat f} =
\frac{n+2}{2 \pi}f$.  

As before, by using the j-function map 
$j({\hat f}) = z$, the energy (\ref{enint1}) can be made finite and is
given by 
\beqa
E = n \, \frac{\pi}{6} \;\;\;.
\label{ener1}
\eeqa
Note that this is half the amount of energy carried by the solutions
with two electric charges that we discussed in section 4.

By expanding $j({\hat f}) = 
e^{2 \pi {\hat f}} + 744 + {\cal O} ( e^{-2 \pi {\hat f}} ) = z$ we see that
$f(\omega) = \frac{1}{n+2} \left( \omega - 744 e^{- \omega} + \dots
 \right) $, which indeed reproduces the correct asymptotic behaviour at
$\omega \rightarrow \infty$.  Thus we see that,  by demanding 
 the asymptotic
behaviour of $f$ to be modified to $f = \frac{2\pi}{n+2} \, j^{-1}(z)$,
the associated energy becomes finite.

We note that the curvature scalar ${\cal R}$ blows up at the 
special point ${\hat f}=1 $.

The solutions discussed above represent one-center
solutions.  They can again be generalised to multi-center solutions via
$j({\hat f}(z)) = P(z)/Q(z)$, where $P(z)$ and $Q(z)$ are polynomials in
$z$ with no common factors.

\section{Eleven dimensional interpretation }

\setcounter{equation}{0}

Heterotic string theory compactified on a seven-torus is related to
M-theory compactified on 
$S_1/{Z_2} \times T_7$ \cite{HorWit}. Hence, our solutions 
should have an eleven dimensional   interpretation in terms of
configurations of intersecting  membranes (M2),   5-branes 
(M5), M-waves and Kaluza--Klein M-monopoles, which are all 
supersymmetric solutions to the low-energy effective action 
of M-theory. These four basic solutions 
all preserve 1/2 of the eleven-dimensional
supersymmetry.  (For a review  see \cite{Gaunt,ts,berg1} 
and references therein.)
 In ten-dimensional heterotic string theory, the basic supersymmetric
 solutions, when reducing from eleven
dimensions,  are a fundamental string (coming from M2), a wave,  
a KK monopole and a NS 5-brane. Each of these ten-dimensional basic
solutions preserve 1/2 of 16 supersymmetry.
Some of the configurations that one gets when considering various 
combinations of these objects, when reduced down to three dimensions,
should  correspond to the solutions constructed in \cite{BBC}. 
The following interpretation emerges when comparing 
 the ten-dimensional metric of each of these 
 four basic building blocks  with
the ten-dimensional ansatz for the vielbein (\ref{viel}):  the wave, when
reduced to three dimensions, gives rise to
 solutions with non-zero $A_\mu^{(1)m}$, which are 
 the gauge fields coming from the reduction of $G^{(10)}_{MN}$; 
the fundamental string gives rise to solutions
with non-zero $A_{\mu m}^{(2)}$,  coming 
from the reduction  of $B^{(10)}_{MN}$; the NS 5-branes give rise 
to solutions with non-zero off-diagonal internal $B_{mn}$ components 
and the Kaluza--Klein monopoles to
solutions with non-zero off-diagonal internal metric $G_{mn}$. 
We will now review the four basic supersymmetric solutions and we will
discuss their reduction to three dimensions in detail.

The eleven-dimensional M-branes, M-waves and M-monopoles are solutions to
the low-energy effective action of M-theory, given by
  $D=11$ supergravity. The
bosonic action contains a metric $g_{MN}$ and a three-form potential
$A_{MNP}$, with field strength $F_{MNPQ}=24\nabla_{[M}A_{NPQ]}:$
\eq
S = \int\sqrt{-g}\bigl\{R-{1\over {12}}F^2 -
  {1\over{432}}\epsilon^{M_1\dots M_{11}}
  F_{M_1\dots M_4}F_{M_5\dots M_8}A_{M_9\dots M_{11}}\bigr\}. 
\en
Supersymmetric solutions to the equations of motion of this action 
can be obtained by looking for backgrounds that admit 32-component
Majorana spinors $\varepsilon$ for which the supersymmetry variation of the
gravitino field $\psi_M$ vanishes.

The M2-brane solution has the form \cite{ds}
\eq
ds_{11}^{2} \,=\, H^{1/3}[\,{1\over H}\,(-dt^2 + dx_1^2 + dx_{11}^2) 
+ (dx_2^2 + \dots dx_9^2)]
\en
with
\eq
F_{t\; 1\; 1\!1\; \alpha} = {c\over 2}{{\del_\alpha H}\over {H^2}},\qquad H =
H(x_2,\dots, x_{9}),\qquad \nabla^2H=0,\qquad c=\pm 1.
\en 
The solution admits Killing spinors  $\varepsilon = H^{-1/6}\,\eta$
with the constant spinor $\eta$ satisfying $\hat{\Gamma}_{0\; 1\; 1\!1}\,\eta =
c\,\eta,$ where $\hat{\Gamma}_{0\dots p}\equiv
\hat{\Gamma_0}\dots\hat{\Gamma}_p$ is the product of $p+1$ distinct Gamma
matrices in an orthonormal frame. Given that $(\hat{\Gamma}_{0\; 1\; 1\!1})^2 
= 1$ and $\Tr\,\hat{\Gamma}_{0\; 1\; 1\!1} = 0$, 
this solution admits 16-component Killing spinors and preserves half of the
supersymmetry. 

The single harmonic function determining the solution depends on the
orthogonal directions to the 2-brane, $\vec{x}=\{x_2,\dots,x_9\}.$
The M2-brane carries electric four-form charge $Q_e$ defined as the
integral of the seven-form $*F$ around a seven-sphere that surrounds the
brane. $c=1$ corresponds to a M2-brane  and
$c=-1$ to an anti-M2-brane.

We now dimensionally reduce the membrane  to ten dimensions to obtain
the fundamental string (NS1) by using that \cite{wit2}
\eq\label{11m}
ds_{11}^2 = e^{2/3\Phi^{(10)}}dx_{11}^2 + e^{-1/3\Phi^{(10)}}ds_{10}^2.
\en
Thus, $e^{2\Phi^{(10)}}=H^{-2}$, and
\eq\label{string}
ds_{10}^2 = H^{-1}(-dt^2+dx_1^2)+dx_2^2 +\dots + dx_9^2.
\en
Here, $ds_{10}^2$ denotes the ten-dimensional line element in the string
frame. 
In order to dimensionally reduce the fundamental string  to three
dimensions, we  take $H$ to be a function of two transverse
coordinates only, 
$H=H(x_8,x_9)$, subject to
$\nabla^2 H = 0$, where $\nabla^2$ now denotes a two-dimensional flat
Laplacian.
We can take the coordinates $x_8$ and $x_9$ to be either cartesian or
cylindrical.  Both coordinate choices, being related by a conformal
transformation, are compatible with $\nabla^2$ being a flat Laplacian.  In
the following, we  take $x_8$ and $x_9$ to be cylindrical coordinates
($\omega=x_8 + i x_9 = \hat{r} + i\hat{\theta}$).

The dimensional reduction of the fundamental string  solution to 
three-dimensions is now obtained by demanding that the 
line element take the form
\eq\label{dr}
ds_{10}^2 =  e^{2\phi^{(3)}}g_{\mu\nu}^{E(3)}dx^\mu dx^\nu +
 G_{mn}^{(7)}dy^mdy^n,
\en
where $G_{mn}^{(7)}$ is the seven-dimensional internal metric with internal
 coordinates $y^m$, and $g_{\mu\nu}^{E(3)}$
is the three-dimensional Einstein metric. Compatibility of the
ten-dimensional action and of the three-dimensional action in 
the Einstein frame  requires that $e^{2\phi^{(3)}} = e^{2\Phi^{(10)}}(\det
G_{mn}^{(7)})^{-1}=H^{-1}$ and 
 the three-dimensional space--time line element in the Einstein 
frame is
\eq\label{3l}
ds^{2} = -dt^2 + H d\omega d\bar{\omega}.
\en
In $D=10$, the fundamental string couples to the antisymmetric 2-tensor
with $B_{tx_1}\propto H^{-1}$, which in $D=3$ yields that 
$A_{tm}^{(2)}={c\over 2}
H^{-1},$ or $F_{t\alpha m}^{(2)}={c\over 2} {{\del_\alpha H}\over{H^2}}$,
where $m=1$.

The solution corresponding to the eleven-dimensional 
M5-brane is of the form \cite{guven}
\eq
ds_{11}^{2} \,=\, H^{2/3}[\,{1\over H}\,(-dt^2 + dx_1^2 +\dots dx_5^2) 
+ (dx_6^2 + \dots + dx_9^2 + dx_{11}^2)]
\en
with
\eq
F_{\alpha_1\dots \alpha_4} = {c\over 2}\epsilon_{\alpha_1\dots\alpha_5} 
\del_{\alpha_5} H,\qquad H =
H(x_6,\dots, x_9,x_{11}),\qquad \nabla^2 H=0,\qquad c=\pm 1,
\en 
where $\epsilon_{\alpha_1\dots\alpha_5}$ is the flat $D=5$ alternating
 symbol.

The solution admits 16-component  Killing spinors $\varepsilon = H^{-1/2}\eta
$ with $\eta$ satisfying $\hat{\Gamma}_{012345}\,\eta = c\eta $.
The M5-brane carries magnetic four-form charge $Q_m$ obtained by
integrating $F$ around a four-sphere that surrounds the M5-brane. Here
again,   $c=\pm 1$ corresponding to a M5- and an
anti-M5-brane respectively.

Using (\ref{11m}), we find that the  M5-brane reduces 
to a NS 5-brane in ten dimensions, with metric
\eq\label{ns}
ds_{10}^2 = -dt^2 + dx_1^2 +\dots + dx_5^2 + H(dx_6^2 + \dots + dx_9^2).
\en
Here  $\, e^{\Phi^{(10)}} = H.$  Setting $\omega = x_8 + ix_9$ and taking
$H=f(\omega)+\bar{f}(\bar{\omega})$, we find that, by using (\ref{dr}),  
the three-dimensional line element 
 is again given by  (\ref{3l}), with $e^{\phi^{(3)}}=1.$ 
Reducing $F_{\alpha_1\alpha_2\alpha_3\alpha_4}$ down to three dimensions
gives rise to $H_{67\alpha}=\del_{\alpha}B_{67}$ with $H_{67\hat{r}}\propto
\del_{\hat{\theta}}H=i\del_{\hat{r}}(f-\bar{f}),\,\,\,
H_{67\hat{\theta}}\propto -\del_{\hat{r}}H=i\del_{\hat{\theta}}(f-\bar{f})$.
So it follows that $B_{67}\propto i(f-\bar{f})$ which is in accordance with
the expressions for the internal $B_{mn}$ field given in sections 4 and 5.

The wave solution in ten dimensions is given by the metric (with
$e^{\phi^{(10)}}=1)$ \cite{brink}
\eq
ds_{10}^2 = -dt^2 + dy_1^2 + (H-1)(dt-dy_1)^2 + (dx_2^2 + \dots + dx_9^2),
\en
or, with $\,y_1 = t + cx_1\,,$
\eq\label{wave}
ds_{10}^2 = 2cdtdx_1  + Hdx_1^2 + (dx_2^2 + \dots + dx_9^2).
\en
This corresponds precisely to  our $n=1$ solutions carrying one electric 
charge,  with a gauge field $\,A_\mu^{(1)a}=A_\mu^{(1)m}e_m^a\,$ coming from
the reduction of the ten-dimensional metric $G_{MN}^{(10)}$.
Indeed, the  ten-dimensional line  element (\ref{viel}) 
used in the reduction is
\eq\label{le}
ds^2_{10} = (e^{2\phi^{(3)}}g_{\mu\nu}^{E(3)} +
A_\mu^{(1)a}\eta_{ab}A_\nu^{(1)b})dx^\mu dx^\nu
+2A_\mu^{(1)a}\eta_{ab}e_n^bdx^\mu dy^n + G_{mn}^{(7)}dy^mdy^n.
\en
Inserting (\ref{cc}) and (\ref{fs}) into (\ref{le}), we recover
 (\ref{wave}), with $c=\eta_{\alpha_1}$,  
where $x_1, x_2\dots x_7$ belong to the seven-dimensional torus $T_7$
and $\omega =x_8+ix_9$.

The Kaluza--Klein monopole in ten dimensions is given by the metric 
\cite{sorkin}
\eq\label{sork}
ds_{10}^2=-dt^2 + Hdx_i^2 + H^{-1}( dz+A_idx_i)^2 + dx_1^2 + \dots +
dx_5^2,\qquad  i=6,8,9,
\en
with $H=H(x_i),\,\,\, F_{ij}=\del_iA_j - \del_j A_i =
c\,\varepsilon_{ijk}\del_k H, \,\,\, e^{2\Phi^{(10)}}=1, \,\,\, c=\pm.$

In five dimensions, the metric reduces to
\eq
ds_5^2 = -dt^2 + Hdx_i^2 + H^{-1}( dz + A_idx_i)^2,
\en
with $e^{2\phi^{(5)}}=1.$

Now, let $H=H(x_8,x_9)$. We can set $A_8=A_9=0.$ Then, $\del_8A_6 =
-c\del_9H,\,\del_9A_6=c\del_8H.$ 
The metric is now
\eqn
ds_5^2 &=& -dt^2 + H(dx_8^2+dx_9^2)+ Hdx_6^2 + H^{-1}( dz 
+ A_6dx_6)^2\no\\
       &=& e^{2\phi^{(3)}}g_{\mu\nu}^{E(3)}dx^\mu dx^\nu + G_{mn}dy^mdy^n,
\enq
with $e^{\phi^{(3)}}=1$ and the off-diagonal internal metric is given by
\eq\label{km}
\ G_{mn}=  \left( \begin{array}{cl}
H+A_6^2 H^{-1} &  A_6H^{-1} \\
 A_6H^{-1} & H^{-1}  
\end{array} \right)\;\;.
\en
We now present the supersymmetric heterotic solutions discussed 
in sections 4 and 5 
as dimensionally reduced solutions corresponding to orthogonally 
intersecting  strings, NS 5-branes, waves and KK monopoles in ten
dimensions.

\subsection{ Solutions carrying two electric charges}

As discussed in section \ref{two}, the three-dimensional space--time line
element in the Einstein frame takes the form 
\eq\label{2n}
ds^2 = -dt^2 + H^{2n}d\omega d\bar{\omega}.\label{linm11}
\en
Consider first the case $n=1$. This solution with diagonal and constant 
internal vielbein $e_m^a$ and $B_{mn}=0$ corresponds to 
a fundamental string and a wave in $D=10$, with the wave travelling along
the string.

In $D=10$, the solution is 
\eq
ds_{10}^2 = (H_s)^{-1}[\,-dt^2 + dy_1^2 + (H_w -1)(dt-dy_1)^2]
+ dx_2^2 +\dots + dx_9^2
\en
or, using (\ref{wave})
\eq\label{ws}
ds_{10}^2 = (H_s)^{-1}[\,2cdtdx_1 + H_w dx_1^2]  
+ dx_2^2 +\dots + dx_9^2.
\en
The ten dimensional dilaton is given by $e^{\Phi^{(10)}}=H_s^{-1}.$
Note that the line element reduces to the wave (\ref{wave}) or string
(\ref{string}) line elements when we respectively set $H_s$ and $H_w$ to 1.
We now set $H_s = H_w = H(x_8,x_9).$
Comparing (\ref{ws}) with (\ref{le}), we find that the internal metric is
constant and that consequently $e^{\phi^{(3)}} = H^{-1}$. Furthermore the gauge
field is given by $A_t^{(1)a}=A_t^{(1)m}e_m^a = c\,H^{-1},$ where $a=1$.
Note that this is in accordance with (\ref{cc1}) and (\ref{fs1}).

Since the wave travels  along the string, they are both characterized by
Killing spinors which obey the same condition $\hat{\Gamma}_{01}\,\eta 
= c_1\,\eta .$ 
This solution thus  preserves 1/2 of 16 supersymmetry.

Note however that, as described in \cite{BBC}, this purely electric
solution can be dualized to a solitonic solution represented by one
off-diagonal entry in the internal metric as well as one non-constant entry
in the internal $B_{mn}$-matrix. These off-diagonal contributions 
play the role of
`magnetic' charge from the ten-dimensional point of view and correspond
respectively to adding a KK monopole and a NS 5-brane, with a common
 5+1-dimensional worldvolume. (This is one of two
possible intersection patterns of a KK monopole and a NS 5-brane in $D=10$ 
\cite{berg1}.)
Taking the 5-brane to be oriented along 
the hyperplane $\{3,4,5,6,7\}$, this configuration is characterized by 
$\hat{\Gamma}_{034567}\,\eta=c_2\eta$ and
$\hat{\Gamma}_{1289}\,\eta=c_3\eta$. The 5-brane and monopole each break
1/2 of 16 supersymmetry. Since, however,  the product of the two Gamma
matrix projection operators gives the ten-dimensional chirality operator
$\Gamma_{11}$, this pair preserves again 1/2 of 16 supersymmetry.
(The ten-dimensional chirality operator, in our notation \cite{BBC}, is
$\Gamma_{11}=\pm\gamma^0\gamma^1\gamma^2\gamma^4\otimes i{\bf I}_8.\,$
$\Gamma_{11}\varepsilon = \varepsilon $ is then equivalent 
to conditions (\ref{cond1})). 
The ten-dimensional line element of this configuration is given by
\eq
ds_{10}^2 = -dt^2 + dx_3^2 + \dots + dx_7^2 + H_5H_{KK}^{-1}(\beta dz +
A_2dx_2)^2 + H_5H_{KK}(dx_2^2 + dx_8^2 + dx_9^2).
\en
Note that this line element reduces to (\ref{ns}) and (\ref{sork}) (up to a
relabelling of the coordinates) when we
respectively set $H_{KK}$ and $H_5$ to 1. The ten-dimensional dilaton is
given by $e^{\Phi^{(10)}}=H_5.$
Setting $H_5=H_{KK}=H(x_8,x_9)$, and comparing with (\ref{dr}), 
yields a constant
three-dimensional dilaton as well as the three-dimensional line element
(\ref{2n}) with $n=1$. 
Furthermore, the  non-constant part of the internal metric $G_{mn}$ 
is found here to be
\eq
\ G_{mn}=  \left( \begin{array}{cl}
\beta^2 & \beta A_2 \\
\beta A_2 & A_2^2 + H^2 
\end{array} \right)\;\;,
\en
where $\beta = \sqrt{G_{11}}$ and where $\del_8 A_2 = -c \del_9 H,\,\del_9
A_2 = c\del_8 H.$
For $H=f+\bar{f}$, it then follows that $A_2 = -ic\;(f-\bar{f}).$
This is indeed the off-diagonal internal metric given in \cite{BBC}.

Consider next the solution with $n=2$.
There is now, in addition to the two electric charges (corresponding to
having a wave and a string), 
 one non-constant off-diagonal entry in the internal metric as
well as one non-constant entry in the $B_{mn}$-matrix. In view of the
discussion above, this amounts to adding a NS 5-brane and a KK
monopole to (\ref{ws}). 
The whole configuration preserves 1/4 of 16 supersymmetry. 
The ten-dimensional line element thus includes a wave, a
fundamental string, a  5-brane and a  monopole. The wave is along
the string and both are in the worldvolume of the NS 5-brane and KK
monopole. The 5-brane and monopole have  common worldvolumes. 
The associated line element is given by
\eqn \label{2n2}
ds_{10}^2 &=& H_s^{-1}[2cdtdx_1+H_wdx_1^2] + dx_2^2 +\dots + dx_5^2 +
                      H_5H_{KK}^{-1}(dx_6 + A_7dx_7)^2\no\\
              &&+ \;H_5H_{KK}(dx_7^2 + dx_8^2 + dx_9^2) \;,
\enq
and the ten-dimensional dilaton is given by 
$e^{\Phi^{(10)}}=H_s^{-1}H_5.$
We now identify $H_s = H_w =H_5 =H_{KK} = H(x_8,x_9)$.
It is easy now to check that the ten-dimensional line element gives rise
to the correct three-dimensional space--time metric (\ref{linm11}) with
$n=2$ as well as to the off-diagonal internal metric given in 
(\ref{gbdyon}). This is done by sending $x_7\longrightarrow {{x_7}\over
  D}$ (where $D = {\sqrt{|\alpha_4 \alpha_{11}|} \over
 \sqrt{|\alpha_2 \alpha_{9}}|}$)  and by relabelling the coordinates.

Consider next the solution with $n=3$.
This solution is obtained from the $n=2$ solution by adding an additional
off-diagonal entry in both the internal metric $G_{mn}$ and in the 
$B_{mn}$-matrix.  This amounts to adding an additional NS 5-brane
and a KK monopole to the line element (\ref{2n2}) in the following way:
\eqn \label{2n3}
ds_{10}^2 &=& H_s^{-1}[2cdtdx_1+H_wdx_1^2] + dx_2^2 + dx_3^2 +
             H_5H_{KK}^{-1}(dx_4 + A_5dx_5)^2 + H_5 H_{KK} dx_5^2 \no\\
           && + \;H_5H_{KK}^{-1}(dx_6 + A_7dx_7)^2
             + H_5H_{KK}dx_7^2 +
H_5^2 H_{KK}^2(dx_8^2 + dx_9^2) \;.
\enq
Note that the 5-branes intersect orthogonally in a 3-brane \cite{ts}
and that the KK monopoles have a common three-dimensional worldvolume
\cite{berg1}. The 
ten-dimensional dilaton is given by 
$e^{\Phi^{(10)}}=H_s^{-1}H_5^2.$ This configuration preserves 1/8 of 16
supersymmetry. 

Finally consider our solution with $n=4$.
 It is obtained by adding  an additional off-diagonal entry 
both in the $G_{mn}$ sector and in the $B_{mn}$ sector. This is achieved by
 adding yet an additional monopole and an
 additional 5-brane.  The resulting
 ten-dimensional line element is given by
\eqn \label{2n4}
ds_{10}^2 &=& H_s^{-1}[2cdtdx_1+H_wdx_1^2] + 
    H_5H_{KK}^{-1}(dx_2 + A_3 dx_3)^2 + H_5 H_{KK} dx_3^2 \no\\
         && + \;    H_5H_{KK}^{-1}(dx_4 + A_5dx_5)^2 + H_5 H_{KK} dx_5^2 
           + H_5H_{KK}^{-1}(dx_6 + A_7dx_7)^2\no\\
           &&   + \; H_5H_{KK}dx_7^2 +
H_5^3 H_{KK}^3(dx_8^2 + dx_9^2) \;.
\enq
The ten-dimensional dilaton is given by 
$e^{\Phi^{(10)}}=H_s^{-1}H_5^3.$
Note that adding 
these two additional objects does not break any further supersymmetry, since
 the Gamma projection operator for the last 5-brane is given by the
 product of the Gamma operators of the wave and the two other 5-branes.
Thus, this configuration also preserves  1/8 of 16
 supersymmetry.

\subsection{Solutions carrying one electric charge}

In this case, the space--time line element in the Einstein frame has the form
\eq
ds^2 = -dt^2 + H^n d\omega d\bar{\omega}.
\en
The $n=1$ solution corresponds in ten dimensions 
to a wave and its reduction is the same as in (\ref{le}) and below.
The wave preserves 1/2 of 16 supersymmetry.

The $n=2$ case can be described in terms of a wave and one NS 5-brane,
since  a non-constant entry in the $B_{mn}$ matrix is added.
The ten-dimensional metric describing this pair is given by
\eq
ds_{10}^2 = 2cdtdx_1 + H_wdx_1^2 + dx_2^2 + dx_3^2 + dx_4^2 + dx_5^2 +
H_5(dx_6^2+\dots + dx_9^2)
\en 
and the ten-dimensional dilaton by $e^{\Phi^{(10)}}=H_5.$

 This solution preserves 1/4 of 16 supersymmetry.
Reducing according to (\ref{le}), we recover our solution given in 
(\ref{gbdyon1}) (up to relabelling of the coordinates) with
$H_w=H_5=H(x_8,x_9)$, 
$\,e^{-2\phi^{(3)}}=H, \,A_t^{(1)a=1}=c/\sqrt{H},\,$ and
$G_{11}=G_{66}=G_{77}=H.$

The $n=3$ ($n=4$) case corresponds in the same way to 
 two (three) orthogonally intersecting 5-branes and a wave, 
where each pair of 5-branes intersects over a 3-brane
\cite{ts}.
For instance, the ten-dimensional line element for the $n=4$ case is given
by
\eq
ds_{10}^2 = 2cdtdx_1 + H_wdx_1^2 + H_5(dx_2^2 + dx_3^2) + H_5(dx_4^2 +
dx_5^2) + H_5(dx_6^2+dx_7^2) + H_5^3(dx_8^2 + dx_9^2)
\en 
and the ten-dimensional dilaton by $e^{\Phi^{(10)}}=H^3,$ where we have
again identified the harmonic functions $H_w=H_5=H(x_8,x_9).$

By using the same argument as in the previous subsection, we
conclude that both these cases preserve 1/8 of 16 supersymmetry.

\section{Conclusions}

We showed that the static supersymmetric solutions constructed in 
\cite{BBC} can be turned into finite energy solutions, which we computed.
The energy of the solutions carrying one electric charge
was found to be given by $E = n \, {\pi \over 6}$, whereas the 
energy of the solutions carrying two electric charges
was found to be  $E = 2n \, {\pi \over 6}$.

The U-duality group of the low-energy heterotic theory is $O(8,24;\IZ)$
\cite{sen1}.  Solutions which are obtained by $O(8,24;\IZ)$ transformations
from the ones discussed above will also have finite energy.  
For instance, the 
compactified heterotic string solutions of \cite{DGHR,DFKR}
will have the same energies as the solutions carrying one electric charge 
discussed in section 5, and similarly for the three-dimensional solutions 
obtained by compactifying a wave and up to  three NS 5-branes and 
 Kaluza--Klein monopoles
intersecting orthogonally in ten dimensions.

The curvature scalar ${\cal R}$ 
associated with the solutions discussed in sections 4 and 5 
is regular everywhere with the
exception of the special point $\phi =0$.  It would be interesting to
understand the physics at this special point 
in moduli space further.

\bigskip
   
{\bf Acknowledgement}  \medskip \newline

We would like to thank I. Bakas, K. Behrndt, R. Khuri, 
D. L\"ust, S. Mahapatra,
T. Mohaupt and J. Schwarz
for valuable discussions.  One of us (G. L. C.) is grateful for the
hospitality at the Institute of Physics of the Humboldt University Berlin,
where part of the work was carried out.

\end{document}